\documentclass{Interspeech}

\interspeechcameraready 

\usepackage{amsmath,graphicx}

\usepackage{amsmath,bm,amssymb,url,cite}
\usepackage{algorithm}
\usepackage{multirow}
\usepackage{algpseudocode}
\usepackage{booktabs}
\usepackage{graphicx}
\usepackage{listings}
\usepackage{xcolor}
\usepackage{url}
\usepackage{hyperref}
\PassOptionsToPackage{hyphens}{url}\usepackage{hyperref}

\newcommand{\hedit}[1]{\textcolor{black}{#1}}
\newcommand{\sedit}[1]{\textcolor{black}{#1}}
\newcommand{\kedit}[1]{\textcolor{black}{#1}}
\newcommand{\kkedit}[1]{\textcolor{black}{#1}}
\newcommand{\kkkedit}[1]{\textcolor{black}{#1}}
\newcommand{\kkkkedit}[1]{\textcolor{black}{#1}}
\newcommand{\kkkkkedit}[1]{\textcolor{black}{#1}}
\newcommand{\keditcamera}[1]{\textcolor{black}{#1}}
\newcommand{\ssedit}[1]{\textcolor{black}{#1}}
\newcommand{\sssedit}[1]{\textcolor{black}{#1}}
\newcommand{\redit}[1]{\textcolor{black}{#1}}

\title{BitTTS: Highly Compact Text-to-Speech \\ Using 1.58-bit Quantization and Weight Indexing}

\author[affiliation={1}]{Masaya}{Kawamura}
\author[affiliation={1}]{Takuya}{Hasumi}
\author[affiliation={1}]{Yuma}{Shirahata}
\author[affiliation={1}]{Ryuichi}{Yamamoto}

\affiliation{}{LY Corporation}{Japan}

\email{kawamura.masaya@lycorp.co.jp, takuya.hasumi@lycorp.co.jp, yuma.shirahata@lycorp.co.jp, ryuichi.yamamoto@lycorp.co.jp}

\usepackage{comment}

\begin{document}

\maketitle
\fontsize{9.0}{10.7}\selectfont
\begin{abstract}
This paper proposes a highly compact, lightweight text-to-speech (TTS) model for on\kedit{-}device applications. To reduce the model size, the proposed model introduces two techniques. First, we introduce quantization\kedit{-}aware training (QAT), which quantizes model parameters during training to as low as 1.58-bit. In this case, most of 32\kedit{-}bit model parameters are quantized to ternary values \{-1, 0, 1\}. Second, we propose a method named weight indexing. In this method, we save \kkkedit{a group of }\sedit{1.58-bit weights} as a single int8 index. This allows for efficient storage of model parameters, even \sedit{on hardware that treats values in units of 8-bit.} \kedit{Experimental results demonstrate that the proposed method achieved \sedit{83}~\% reduction in model size, \sssedit{while outperforming the baseline of similar model size without quantization in synthesis quality.}} 

\end{abstract}

\section{Introduction}
\label{sec:intro}
Many high-quality neural text-to-speech (TTS) models have been extensively researched~\cite{zen2013tts,ning2019review,tan2021survey,andreas2023overview, MEHRISH2023101869}. With the advancement of these TTS models, they are increasingly being integrated into mobile applications, such as car navigation systems and conversational bots, among others. 
\sedit{In these applications, TTS models are typically required to generate speech quickly on low-end devices with limited memory and storage. 
Therefore, research on 1) fast inference with limited computational resources and 2) model size compression\keditcamera{\footnote{\keditcamera{
In addition to the storage usage, model size compression is also beneficial for TTS application users to save mobile data usage.
}}} \keditcamera{is} crucial for ensuring that TTS technology can be effectively used in many real-world applications.}

\sedit{To address the former, various lightweight TTS models have been proposed and demonstrated promising results~\cite{luo2021lightspeech, Chevi2023nixtts, okamoto2024convnext, kawamura2023mbvits, vainer2020speedy, ping2018ClariNet}. The methods include multi-band generation~\cite{kawamura2023mbvits} and convolution-based lightweight blocks~\cite{okamoto2024convnext}, as well as other approaches.
On the other hand, for the model size compression of lightweight TTS models, research in this area is still in its early stages. Although some studies have tackled this through knowledge distillation~\cite{Chevi2023nixtts, vainer2020speedy, ping2018ClariNet} or neural architecture search~\cite{luo2021lightspeech, Zoph_2018_CVPR, NEURIPS2018_933670f1}, there remains room for further investigation.}

\sedit{\kedit{O}ne} approach to reduce the model size of neural network models is quantization\kedit{~\cite{Gholami2021quan, Nagel2021nnq}}. Quantization reduces the model size by representing weights with low precision. \kedit{According to \cite{Gholami2021quan, Nagel2021nnq}, \sedit{this}} can be roughly categorized into post-training quantization (PTQ) and quantization-aware training (QAT). 
PTQ involves quantizing a pre-trained model without further training, which can lead to significant degradation in model performance due to the abrupt reduction in numerical precision. In contrast, QAT incorporates quantization into the training process, allowing the model to adapt to the lower precision weights. This enables QAT to mitigate the accuracy loss typically associated with quantization, as confirmed by previous studies in fields such as computer vision~\kedit{\cite{Jacob_2018_CVPR, Gholami2021quan}}.

In addition, while many studies have investigated 8-bit or 4-bit quantization of neural network models~\cite{ding20224bit, zhen2022sub8bit}, some recent studies report that extremely low precision such as 1 or 1.58-bit still works well in tasks such as \kedit{language} modeling\kedit{~\cite{wang2023bitnet, ma2024era}} \kedit{and computer vision}~\cite{xnor2016rastegari, Bulat2019XNORNet, QIN2020107281, Courbariaux2016binary, Xia2023basic}.
Despite the success of QAT in many fields, it has not been extensively studied in the context of lightweight TTS.

In this paper, we propose a lightweight TTS model with 1.58-bit quantization to reduce the size of TTS models for on-device applications. Since QAT has not been investigated in the TTS field so far, we also evaluate 4-bit variants to explore the trade-offs between model size and synthesized speech quality. In the 1.58-bit quantization setting, the model parameters are quantized to ternary values $\{-1, 0, 1\}$. This setting is expected to significantly reduce the size of each model \kedit{weight}, which is otherwise represented by 32\kedit{-}bit (float32). However, it is typically difficult to save the quantized weights with \kedit{their} original amount of information (i.e.\kedit{,} save each \kedit{weight} as 1.58\kedit{-}bit) due to hardware limitations, which handle all weights as 8-bit or 16-bit units. To overcome this limitation, we propose a method named \textit{weight indexing}, which saves the model weights as indices of a predefined set of \textit{weight patterns}. 
Specifically, since the total number of 5 weight combinations is at most $3^5 = 243$, every 5 weights are stored as a single 8-bit integer, which can represent $2^8 =256$ patterns\footnote{\sedit{We use 5 here because 243 patterns can be efficiently represented by the 256 patterns supported by an 8-bit integer. In other words, we store $1.58\times5=7.9$ bit into an 8\kedit{-}bit integer.}}. This enables saving the weights at a size close to the original amount of information.
\kkkedit{We call the TTS model with quantization and weight indexing as \textit{BitTTS}.}

Experimental results on lightweight TTS models showed that the proposed method with weight indexing can reduce the model size up to \kedit{83}~\% when both the acoustic model and the vocoder are quantized. \sssedit{In terms of the synthesis quality, the model outperformed a baseline model of similar model size without quantization.}
We also found that avoiding the quantization of the vocoder significantly improves the synthesized speech quality. Even in this case, the model size was still reduced by about \kedit{70}~\%. 
\kkkedit{Audio samples are available on our demo page\footnote{\url{https://masayakawamura.github.io/bittts-demo/}}.}

\begin{figure*}[tb]
\centering 
\includegraphics[scale=0.44]{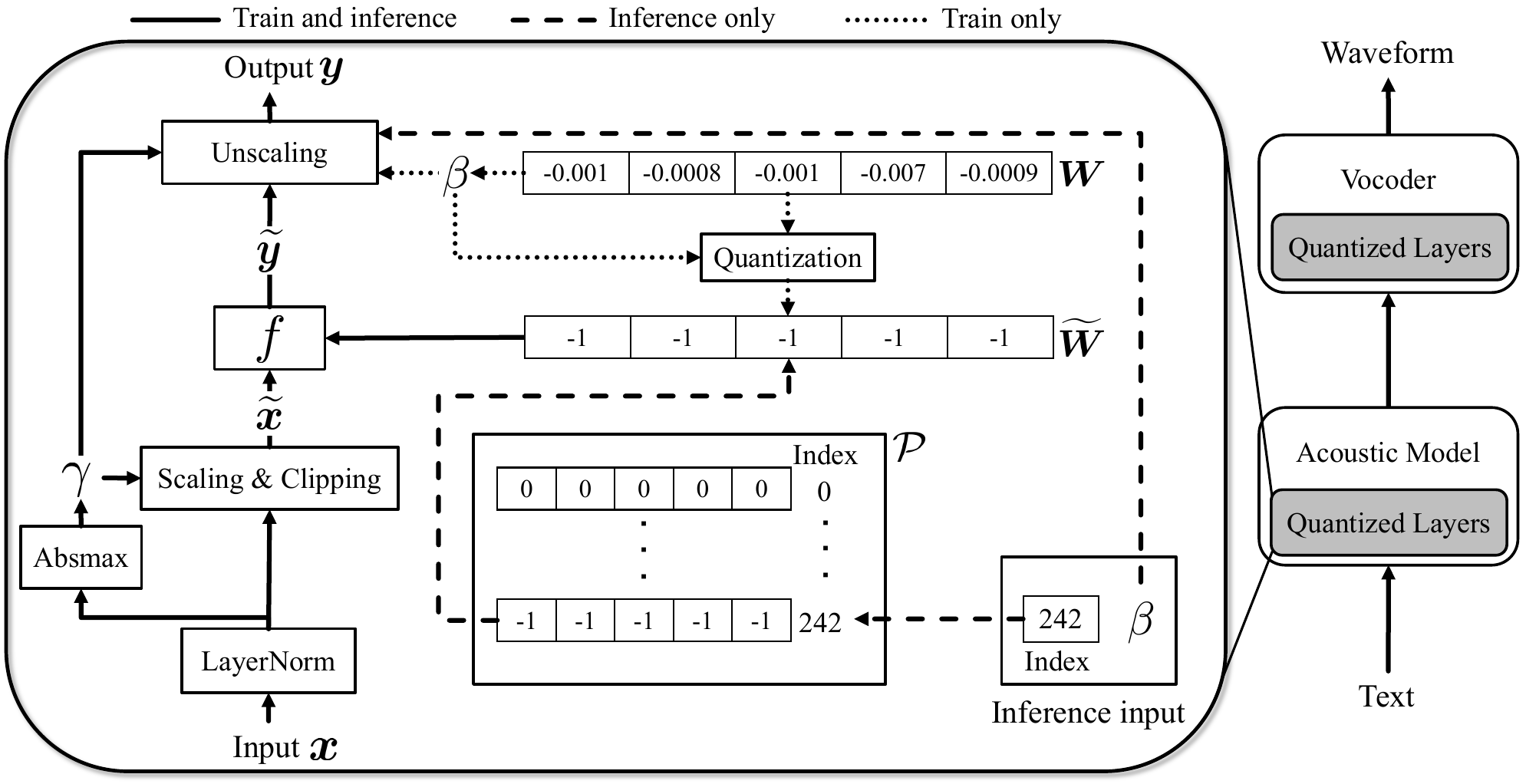}
\caption{
\kkkkedit{
BitTTS quantizes the weight of TTS. The proposed method reduces the model size by storing indices instead of weights.}
}
\label{fig: proposed}
\vspace{-3mm} 
\end{figure*}
\if 0
QAT is a technique designed to address the limitations of PTQ by incorporating quantization effects directly into the training process. While PTQ converts a pre-trained floating-point model into a fixed-point model without retraining, it often struggles with significant accuracy drops, especially when targeting low-bit quantization. Conversely, QAT simulates quantization noise during training, allowing the model to adapt to these effects and thus maintain higher accuracy.
\else
\if

\section{\kedit{Method}}

\subsection{Quantization-\kkkkedit{a}ware \kkkkedit{t}raining}
\label{sec:qat}

\hedit{QAT~\cite{Jacob_2018_CVPR} is a technique to quantize models dynamically during training. This quantization approach is widely used in not only natural language processing~\cite{wang2023bitnet,ma2024era} but also speech recognition~\cite{ding20224bit,rybakov20232bit} due to its higher accuracy~\cite{Jacob_2018_CVPR,Gholami2021quan} than the static quantization such as PTQ.}
\hedit{In this paper, we applied \kedit{a} popular QAT method called fake QAT~\cite{Jacob_2018_CVPR}.}

\subsection{\kedit{Quantization \kkkkedit{f}ormulation}}\label{sec: bitconv}

\kedit{
The weights of a neural network layer are denoted as \( \kedit{ \bm{W}} = \{w_1, w_2, \dots, w_N\}\), \sedit{where $w_i$ denotes \kkkedit{$i$}-th scalar weight\kkkedit{, and $N$ represents the total number of weights}.}
}
Inspired by the BitNet \kedit{b1.58 method~\cite{ma2024era}}, 
we scale the weights $\bm{W}$ by their average absolute values and then round each value to the nearest integer. \kedit{
When quantizing $\bm{W}$ to $b$-bit, the following equations are used: 
}
\begin{align}\label{eq: quan_w2}
&\widetilde{\kedit{\bm{W}}} = \text{Round}\left(\text{Clip}\left( \frac{\kedit{\bm{W}}}{\beta + \epsilon}, -q, q-1 \right)\right), \\
&q = 2^{(b-1)}, \\
&\text{Clip}(u, l, h) = \text{max}(l, \text{min}(h, u)).
\end{align}
\kedit{
Specifically, when quantizing $\bm{W}$ to 1.58-bit, it is quantized to ternary values $\{-1, 0, 1\}$:
}
\kedit{
\begin{align}\label{eq: quan_w}
\widetilde{\kedit{\bm{W}}} &= \text{Round}\left(\text{Clip}\left( \frac{\kedit{\bm{W}}}{\beta + \epsilon}, -1, 1 \right)\right),
\end{align}
}%
where $\epsilon$ is a small floating-point number that prevents division by zero. $\beta$ is calculated as follows:

\kedit{
\begin{align}
\beta=\frac{1}{N} \sum_{w\in \kedit{\bm{W}}}|w|.
\end{align}
}

\kedit{
In the case of convolutional layers, the weights \( \kedit{\bm{W}} \) depend on the number of output channels \( c_{\text{out}} \), the number of input channels \( c_{\text{in}} \), and the kernel size \( K \). The weights are represented as \( \kedit{\bm{W}} \in \mathbb{R}^{c_{\text{out}} \kkkedit{\times}c_{\text{in}} \kkkedit{\times}K} \), and \( \beta \) is calculated as follows:
}
\kedit{
\begin{align}
\beta = \frac{1}{c_{\text{out}} c_{\text{in}} K} \sum_{i=1}^{c_{\text{out}}} \sum_{j=1}^{c_{\text{in}}} \sum_{k=1}^{K} |\kedit{\bm{W}}_{i, j, k}|,
\end{align}
}%
where \( \kedit{\bm{W}}_{i, j, k}\) represents the weight corresponding to the \( i \)-th output channel, the \( j \)-th input channel and the \(k\)-th kernel index, respectively.

\kkkedit{
Layer normalization~\cite{layernorm2016ba} is applied to the input tensor $\bm{x}$. After the layer normalization, the
}
input $\bm{x}$ is scaled to the range $[-Q_p, Q_p]$ where $Q_p = 2^{p - 1} $, with $p$ representing $p$-bit precision. The scaling and clipping are performed as follows:
\begin{align}
\widetilde{\kedit{\bm{x}}} &= \text{Clip}\left(\frac{\kedit{\bm{x}}Q_p}{\gamma}, -Q_p + \epsilon, Q_p - \epsilon \right), \\ \gamma &= ||\kedit{\bm{x}}||_{\infty}. \label{eq: clip}
\end{align}
\kedit{Here, \(\gamma\) is the infinity norm of the input tensor \( \kedit{\bm{x}} \), representing the maximum absolute value within \( \kedit{\bm{x}} \).}

\kedit{
The output can be obtained from the quantized weights \(\widetilde{\kedit{\bm{W}}}\) and input \(\widetilde{\kedit{\bm{x}}}\). Let \( f \) represent the neural network operation, then it can be written as follows:
}
\begin{align}
\widetilde{\kedit{\bm{y}}} = f(\widetilde{\kedit{\bm{W}}}, \widetilde{\kedit{\bm{x}}}).
\end{align}
\kedit{
For example, in the case of a convolutional layer, \( f \) represents the convolution operation.} 
To restore the original scale, the output \( \widetilde{\kedit{\bm{y}}} \) is rescaled by multiplying it by the scale factors \( \gamma \) and \( \beta \), and dividing by the quantization range \( Q_p \). The rescaled output \( \kedit{\bm{y}} \) can be expressed as:
\begin{align}
\kedit{\bm{y}} = \frac{\widetilde{\kedit{\bm{y}}} \gamma  \beta }{Q_p}\keditcamera{.}
\end{align}
Following BitNet \kkkedit{b1.58~\cite{ma2024era}}, we do not use the bias. Due to the non-differentiable functions such as clipping and rounding, we apply the straight-through estimator~\cite{Bengio2013ste} during backpropagation to approximate the gradient.  

\subsection{Weight \kkkkedit{i}ndexing}\label{sec: indexing}
\if 0
Even if a model with ternary values \{-1, 0, 1\} is implemented, it does not effectively reduce the model size if the hardware can only handle 8-bit or 4-bit units.
\else
\hedit{If a model \redit{weight} is quantized with ternary values \{-1, 0, 1\}, each value can be ideally represented by $\log_{2}3\approx 1.58$-bit. However, since most deep learning frameworks or hardware 
\sedit{typically handle values in units of 8-bit or 16-bit,}
quantization by ternary values does not effectively reduce the model size. \sedit{This is because each 1.58-bit weight is saved as 8-bit or 16-bit value.}}
\fi
To efficiently 
\sedit{save weights,}
\hedit{we store the integer indices of a predefined set of weight patterns $\mathcal{P} = \{(0,\ldots,0,0)^{\mathsf{T}},(0,\ldots,0,1)^{\mathsf{T}},\ldots,(-1,\ldots,-1,-1)^{\mathsf{T}}\}$ instead of quantized weights $\widetilde{\bm{W}}$ itself.}
\fi

\if 0
Algorithm~\ref{alg:weight_indexing} outlines the steps to generate ternary patterns \(\{-1, 0, 1\}\) for a given kernel size \(K\) or a threshold \(K_\text{threshold}\). The reason for setting the threshold \(K_\text{threshold}\) is that when the kernel size is large, the number of ternary patterns for the kernel size becomes enormous, leading to a large number of bits required to store the indices. If the kernel size exceeds this threshold, the quantized weights $\widetilde{W}$ are split into smaller segments. Each segment is then matched with the patterns in $\mathcal{P}$, and the corresponding indices are stored.
\else
\hedit{Let us assume that the quantized weight $\widetilde{\bm{W}}$ can be represented as a flattened vector $\widetilde{\bm{v}}\in\mathbb{R}^{L}$ to generalize.
Algorithm~\kedit{\ref{alg:generalized_weight_indexing}} outlines the steps to generate a list of weight patterns for a flattened vector $\widetilde{\bm{v}}$.
In this algorithm, we first define a block size $L^*$, by which the whole $L$ weights are divided into segments of length $L^*$. Then, every $L^*$ \kkkkkedit{weight} \kkkkkedit{is} mapped to a virtual lookup table representing weight patterns $\mathcal{P}$ with length $3^{L^*}$. Finally, the list of weight patterns is saved.}
\fi

\hedit{In practice, to represent weight patterns as 8-bit integers, we choose $L^{*}=5$. 
If $L^{*}=5$, $3^{L^{*}}\approx 2^{8}$ holds, which leads to the effective use of the bit length in 8-bit. 
\kedit{
For example, in the case of 1D convolution with $c_{\mathrm{in}}=c_{\mathrm{out}}=256$ and kernel size $K=5$, the weight can be ideally saved as $63.4\mathrm{kB}$\footnote{$(c_{\mathrm{in}}\times c_{\mathrm{out}} \times L^{*} \times \log_{2}3) / (8 \times 2^{10})=63.4$} when quantized to 1.58-bit.
}
\kedit{
However, if the hardware only supports 8-bit as the minimum precision, the weight is saved as $320$kB\footnote{$(c_{\mathrm{in}}\times c_{\mathrm{out}} \times L^{*} \times 8) / (8 \times 2^{10})=320$}.}
By using our algorithm, the quantized weight is saved as $64.0$kB\footnote{$(c_{\mathrm{in}}\times c_{\mathrm{out}} \times 8) / (8 \times 2^{10})=64.0$}, which shows the significant model size reduction.} 

\kedit{Figure~\ref{fig: proposed} shows the quantization of the TTS model using the proposed method.} 
The stored indices are used to quickly reconstruct the quantized weights $\widetilde{\kedit{\bm{W}}}$ from the predefined patterns $\mathcal{P}$. This involves retrieving the corresponding pattern for each index and concatenating the segments to form the complete weight tensor. 

\begin{algorithm}[t]
\caption{Weight Indexing}
\label{alg:generalized_weight_indexing}
\begin{algorithmic}[1]
\State \textbf{Input:} $\widetilde{\bm{v}}=(\widetilde{v}_{1},\ldots,\widetilde{v}_{L})^{\mathsf{T}} \in \{-1, 0, 1\}^{L}$, block size $L^*$
\State \textbf{Output:} List of indices of weight pattern $\mathcal{P}$
\State Initialize $\text{indices} = []$
\For {each block in $\{1, 2, \ldots, \lceil L / L^* \rceil\}$}
    \State Initialize $n = 0$
    \For {each $l$ in $\{ \min(\text{block} \times L^*, L), \ldots, (\text{block} - 1) \times L^* + 1 \}$}
        \State $n \leftarrow n \times 3$
        \If{$\widetilde{v}_{l}=1$}
            \State $n \leftarrow n + 1$
        \ElsIf{$\widetilde{v}_{l}=-1$}
            \State $n \leftarrow n + 2$
        \EndIf
    \EndFor
    \State Append $n$ to $\text{indices}$
\EndFor
\State \textbf{Return:} List of indices
\end{algorithmic}
\end{algorithm}

\section{Experiments}

\begin{table*}[tb]
\caption{
Comparison of model size and naturalness MOS with 95~\% confidence intervals and average RTF on an Apple M1 Pro.} 
\label{tab: result1}
\centering
\scalebox{1.0}{
    \begin{tabular}{ccccccc}\toprule
        Method & \multicolumn{2}{c}{Quantization} & Weight Indexing & Model Size [MB] & MOS & RTF \\ 
            & Acoustic Model & Vocoder & & & & \\\midrule
        32-bit & - & - & - & 25.66 & $3.75 \pm 0.09$ & 0.042 \\
        32-bit (small model) & - & - & - & 12.78 & $1.20 \pm 0.04$ & 0.019\\
        4-bit & \checkmark &  & - & 11.40 & $3.34 \pm 0.10$ & 0.047  \\
        4-bit &  & \checkmark & - & 23.12 & $2.57 \pm 0.10$ &  0.062 \\
        4-bit & \checkmark & \checkmark & - & 8.87 & $2.96 \pm 0.11$ & 0.068 \\\cmidrule{1-7}
        \multirow{2}{*}{1.58-bit} & \checkmark &  &  & 11.40 & \multirow{2}{*}{$3.30 \pm 0.10$} & \multirow{2}{*}{0.040}  \\
         & \checkmark &  & \checkmark & 7.60 & &  \\\cmidrule{1-7}
        \multirow{2}{*}{1.58-bit} &  & \checkmark &  & 23.12  & \multirow{2}{*}{$3.18 \pm 0.11$} & \multirow{2}{*}{0.059}  \\ 
         &  & \checkmark & \checkmark & 22.45 & &  \\\cmidrule{1-7}
        \multirow{2}{*}{1.58-bit} & \checkmark & \checkmark &  &8.87  & \multirow{2}{*}{$3.09 \pm 0.11$} & \multirow{2}{*}{0.064} \\
         & \checkmark & \checkmark & \checkmark & 4.39 &  &  \\\midrule
        Ground truth & - & - & - & - & $4.24 \pm 0.09$ & - \\\bottomrule
    \end{tabular}
    }
\vspace{-3mm}
\end{table*}

\subsection{Experimental \kkkkedit{c}onditions}\label{sec:exp_cond}
\sedit{We conducted experiments to evaluate the effectiveness of quantization in TTS and the proposed methods.}
We used LibriTTS-R~\cite{koizumi2023librittsr}, a high-quality multi-speaker corpus for experiments.
This corpus comprises 585 hours of speech data at a 24~kHz sampling rate from 2,456 English speakers. We divided the corpus into a training set (475.61~hours), a validation set (0.17~hours), and a test set (51.92~hours). To train the duration model in TTS models, we extracted phone duration by Montreal forced aligner~\cite{mcauliffe2017montreal}.

\kkedit{
    To evaluate the effectiveness of the proposed method, we conducted experiments using a TTS model. 
    As the base TTS architecture, we adopted JETS~\cite{lim2022jets} \kkkedit{for its simple architecture}. \ssedit{However, since the transformer blocks~\kkkedit{~\cite{vaswani2017attention}} in the encoder and the decoder of JETS are computationally expensive, these blocks were substituted with 4 convolutional layers of kernel size 5.}}
\kedit{
The number of channels was set to 256 for the encoder and 192 for the decoder.}
The vocoder used is HiFi-GAN\kedit{~\kedit{\cite{kong2020hifi}}}. The initial channel in HiFi-GAN was set to 128. 
\ssedit{Additionally, to confirm the effectiveness of the quantization for model size compression, we prepared another 32-bit baseline model. In this model, the model size was reduced using small channel sizes instead of quantization.}
\kkkedit{For this model, the channel of the encoder and decoder in the acoustic model and the initial channel of HiFi-GAN were set to 32.}
\kedit{\redit{We omit the energy predictor from these models for simplicity}. Additionally, to improve naturalness, we added a \kedit{mixture density network~\cite{Bishop94mixturedensity}} layer to the duration predictor~\cite{du2021phone}.}

\kkkedit{
We used an NVIDIA A100 GPU to train all the models. We utilized AdamW optimizer~\cite{Ilya2019adamw} with $\beta_1=0.8$, $\beta_2=0.99$. The initial learning rate was $2 \times 10^{-4}$. The learning rate decay was scheduled using the \kedit{e}xponential learning rate scheduler~\cite{Li2019explr} with a decay factor of 0.9973. We used a dynamic batch size with an average of \kedit{15} to create a minibatch\kedit{~\cite{hayashi2020esp}}.
All models were trained for 2000~K steps.
}

\subsection{\kkkedit{Quantization \kkkkedit{d}etails}}
\sedit{For quantization, considering that more than 90 \% of the model weights in the baseline TTS models belong to 1D convolutional layers, we applied quantization to only 1D convolutional layers in baseline TTS models for simplicity. Quantized weights in convolutional layers were stored as int8, while other weights were stored as float32.}
\sedit{In the case of 4-bit quantization, the quantization method follows the approach in Section~\ref{sec: bitconv}, with the clipping range in Equation~\ref{eq: quan_w} set to \(- (2^{(4 - 1)}) = -8\) to \(2^{(4 - 1)} - 1 = 7\). Additionally, \( p \) was set to 8. In the case of 1.58-bit, the quantization method follows the approach in Section~\ref{sec: bitconv}, and the weight indexing from Section~\ref{sec: indexing} was applied.}
\kedit{Note that when quantizing the TTS vocoder to 4-bit \sedit{or} 1.58-bit, the convolutional layer closest to the waveform output was excluded from quantization. This is because preliminary experiments showed that quantizing \sedit{this layer} significantly degrades the quality of the synthesized speech.}
Additionally, to investigate the impact of quantizing different modules, we conducted experiments in three settings: quantization of 1) \kkkedit{both the acoustic model and vocoder, 2) only the acoustic model, and 3) only the vocoder.}

\subsection{\kkkedit{Evaluation}}

\subsubsection{\kkkkedit{Naturalness of synthesized speech}}
\kkkedit{
To evaluate the quality of generated samples, we conducted a subjective listening test using five-point naturalness mean opinion score (MOS).
We randomly selected 30 utterances from the test data and synthesized speech \sedit{samples}. Then, \kkedit{15} raters listened to \sedit{the samples} for evaluation.}
The experimental results are shown in Table~\ref{tab: result1}.
\kkkkedit{The results indicate that the 32-bit small model, created by naively reducing the size of the 32-bit TTS model, significantly decreases the MOS compared with the base 32-bit TTS model.}
In contrast, the 4-bit and 1.58-bit TTS models achieve higher MOS compared to the smaller 32-bit model. 
\kkkedit{This result indicates that quantization is effective for creating small TTS models without significantly degrading quality.}

\ssedit{When comparing the 4-bit and 1.58-bit TTS models, the 1.58-bit models achieve comparable or even higher MOS despite the reduced model size.}
\ssedit{Particularly, when the vocoder is quantized, there is a significant MOS gap.
\kkkkedit{
One possible reason could be the instability in vocoder training due to quantization, which might have resulted in the 4-bit vocoder not learning effectively in this experiment.}
}

\ssedit{Comparing the modules to be quantized, the quantization of the vocoder results in lower MOS than that of the acoustic model. This could be because the vocoder, which generates the waveform, is more sensitive to quality degradation when quantized than the acoustic model. In addition, quantizing the acoustic model resulted in a greater reduction in the model size. This is because, in the TTS model used in this experiment, the majority of the model size is attributed to the acoustic model. Given these results, quantizing the acoustic model would be one of the best options in terms of the model size and the synthesis quality.}

Furthermore, the 1.58-bit TTS models with weight indexing achieve a smaller model size than the 4-bit models and 1.58-bit TTS without weight indexing, reducing the model size by up to 83~\%\footnote{$(25.66 - 4.39) / 25.66 \times 100$} compared to the base 32-bit TTS model. 
This result demonstrates that weight indexing is highly effective in reducing model size. \sssedit{When the vocoder \kkkkkedit{was} not quantized for better synthesis quality, the model size was still reduced to 70~\%.}

\subsubsection{\kkkkedit{Inference speed}}
\kkkedit{
We measured the average real-time factor (RTF) on an Apple M1 Pro \sedit{CPU} using \kkedit{100} randomly sampled utterances from the test set. For the method using weight indexing, we first restored the weights from the index and then measured the RTF.
}

\kkkedit{
Regarding the RTF, the small 32-bit model marked the best score. 
\kkkkedit{This is because the complexity of the 32-bit small TTS model is the lowest by removing a significant portion of the parameters from the base 32-bit TTS model.}
Quantized models tend to be slightly slow due to additional layers and operations introduced by the quantization method described in Section~\ref{sec: bitconv}.
When the vocoder is quantized, this effect becomes more pronounced due to the vocoder's high computational cost.
In contrast, the processing of the acoustic model is \ssedit{more lightweight} than the vocoder, resulting in minimal impact on RTF when quantized. 
Therefore, if inference speed is a priority, quantizing only the acoustic model \ssedit{would be the best choice.} 
}

\subsubsection{\kkkkedit{Analysis of weight patterns}}
Figure~\ref{fig: pattern_histogram} shows the distribution of indices after applying weight indexing. The distribution is calculated from the 1.58-bit quantized TTS model with both the acoustic model and vocoder quantized. The result indicates that the indices are biased towards specific values and are not uniformly distributed. In particular, indices 0, 121, and 242 appear more frequently, corresponding to weight values of 0, 1, and -1 repeated \(L^*\) times, respectively. Future work will explore methods to further reduce the model size by leveraging this bias in the occurrence frequency of indices\footnote{After applying Huffman coding~\cite{huffman52} to the model, which quantized both the acoustic model and vocoder to 1.58\kkkedit{-bit} and using weight indexing, the model size was reduced from 4.39~MB to 4.35~MB.}.

\begin{figure}[tb]
    \centering 
    \includegraphics[scale=0.40]{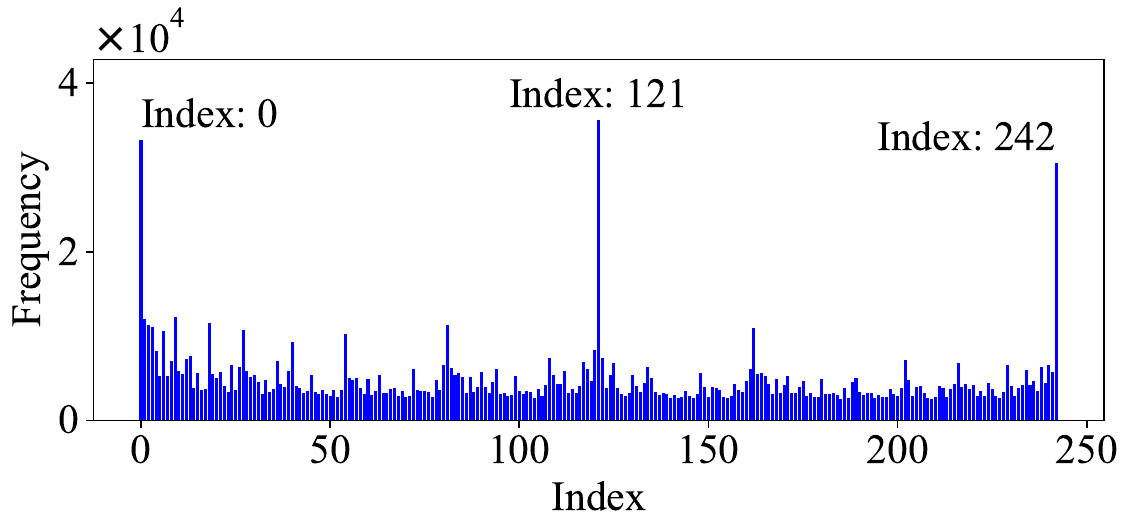}
    \caption{
    Frequency of indices when applying weight indexing to the 1.58-bit TTS model.
    }
    \label{fig: pattern_histogram}
    \vspace{-3mm}
\end{figure}

\section{Conclusions}

\kedit{
In this paper, we proposed a highly compact TTS model with 1.58-bit quantization for on-device applications. Additionally, we introduced a method called weight indexing to further reduce model size. Since QAT has not been explored in the TTS field, we also evaluated the trade-off between model size and synthesized speech quality. 
\kkkedit{
Experimental results showed that our proposed method could reduce the model size by over 80~\% while achieving a naturalness MOS comparable to or higher than that of the 4-bit TTS models.
}
Furthermore, we found that quantization of the vocoder significantly affects the quality of synthesized speech in TTS models. 
Future work includes improving the quality of TTS using the proposed method and applying weight indexing to other models.
}

\vfill\pagebreak

\bibliographystyle{IEEEtran}
\bibliography{ref}

\begin{thebibliography}{10}
\providecommand{\url}[1]{#1}
\csname url@samestyle\endcsname
\providecommand{\newblock}{\relax}
\providecommand{\bibinfo}[2]{#2}
\providecommand{\BIBentrySTDinterwordspacing}{\spaceskip=0pt\relax}
\providecommand{\BIBentryALTinterwordstretchfactor}{4}
\providecommand{\BIBentryALTinterwordspacing}{\spaceskip=\fontdimen2\font plus
\BIBentryALTinterwordstretchfactor\fontdimen3\font minus \fontdimen4\font\relax}
\providecommand{\BIBforeignlanguage}[2]{{%
\expandafter\ifx\csname l@#1\endcsname\relax
\typeout{** WARNING: IEEEtran.bst: No hyphenation pattern has been}%
\typeout{** loaded for the language `#1'. Using the pattern for}%
\typeout{** the default language instead.}%
\else
\language=\csname l@#1\endcsname
\fi
#2}}
\providecommand{\BIBdecl}{\relax}
\BIBdecl

\bibitem{zen2013tts}
{H. Zen}, {A. Senior}, and {M. Schuster}, ``Statistical parametric speech synthesis using deep neural networks,'' in \emph{Proc. ICASSP}, 2013, pp. 7962--7966.

\bibitem{ning2019review}
{Y. Ning}, {S. He}, {Z. Wu}, {C. Xing}, and {L.-J. Zhang}, ``A review of deep learning based speech synthesis,'' \emph{Appl. Sci.}, vol.~9, no.~19, 2019.

\bibitem{tan2021survey}
{X. Tan}, {T. Qin}, {F. Soong}, and {T.-Y. Liu}, ``A survey on neural speech synthesis,'' \emph{arXiv preprint arXiv:2106.15561}, 2021.

\bibitem{andreas2023overview}
{A. Triantafyllopoulos}, {B. W. Schuller}, {G. I\.{y}men}, {M. Sezgin}, {X. He}, {Z. Yang}, {P. Tzirakis}, {S. Liu}, {S. Mertes}, {E. André}, {R. Fu}, and {J. Tao}, ``An overview of affective speech synthesis and conversion in the deep learning era,'' \emph{Proc. of the IEEE}, vol. 111, no.~10, pp. 1355--1381, 2023.

\bibitem{MEHRISH2023101869}
{A. Mehrish}, {N. Majumder}, {R. Bharadwaj}, {R. Mihalcea}, and {S. Poria}, ``A review of deep learning techniques for speech processing,'' \emph{Information Fusion}, vol.~99, p. 101869, 2023.

\bibitem{luo2021lightspeech}
{R. Luo}, {X. Tan}, {R. Wang}, {T. Qin}, {J. Li}, {S. Zhao}, {E. Chen}, and {T.-Y. Liu}, ``Lightspeech: Lightweight and fast text to speech with neural architecture search,'' in \emph{Proc. ICASSP}, 2021, pp. 5699--5703.

\bibitem{Chevi2023nixtts}
{R. Chevi}, {R. E. Prasojo}, {A. F. Aji}, {A. Tjandra}, and {S. Sakti}, ``{NIX-TTS}: Lightweight and end-to-end text-to-speech via module-wise distillation,'' in \emph{Proc. SLT}, 2023, pp. 970--976.

\bibitem{okamoto2024convnext}
T.~Okamoto, Y.~Ohtani, T.~Toda, and H.~Kawai, ``{ConvNeXt-TTS and ConvNeXt-VC}: {ConvNeXt}-based fast end-to-end sequence-to-sequence text-to-speech and voice conversion,'' in \emph{Proc. ICASSP}, 2024, pp. 12\,456--12\,460.

\bibitem{kawamura2023mbvits}
M.~Kawamura, Y.~Shirahata, R.~Yamamoto, and K.~Tachibana, ``Lightweight and high-fidelity end-to-end text-to-speech with multi-band generation and inverse short-time fourier transform,'' in \emph{Proc. ICASSP}, 2023, pp. 1--5.

\bibitem{vainer2020speedy}
J.~Vainer and O.~Dušek, ``{SpeedySpeech}: Efficient neural speech synthesis,'' in \emph{Proc. Interspeech}, 2020, pp. 3575--3579.

\bibitem{ping2018ClariNet}
{W. Ping}, {K. Peng}, and {J. Chen}, ``{ClariNet}: Parallel wave generation in end-to-end text-to-speech,'' in \emph{Proc. ICLR}, 2019.

\bibitem{Zoph_2018_CVPR}
B.~Zoph, V.~Vasudevan, J.~Shlens, and Q.~V. Le, ``Learning transferable architectures for scalable image recognition,'' in \emph{Proc. CVPR}, 2018.

\bibitem{NEURIPS2018_933670f1}
R.~Luo, F.~Tian, T.~Qin, E.~Chen, and T.-Y. Liu, ``Neural architecture optimization,'' in \emph{Proc. NeurIPS}, vol.~31, 2018.

\bibitem{Gholami2021quan}
A.~Gholami, S.~Kim, Z.~Dong, Z.~Yao, M.~W. Mahoney, and K.~Keutzer, ``A survey of quantization methods for efficient neural network inference,'' \emph{arXiv preprint arXiv:2103.13630}, 2021.

\bibitem{Nagel2021nnq}
{M. Nagel}, {M. Fournarakis}, {R. A. Amjad}, {Y. Bondarenko}, {M. van Baalen}, and {T. Blankevoort}, ``A white paper on neural network quantization,'' \emph{arXiv preprint arXiv:2106.08295}, 2021.

\bibitem{Jacob_2018_CVPR}
B.~Jacob, S.~Kligys, B.~Chen, M.~Zhu, M.~Tang, A.~Howard, H.~Adam, and D.~Kalenichenko, ``Quantization and training of neural networks for efficient integer-arithmetic-only inference,'' in \emph{Proc. CVPR}, 2018, pp. 2704--2713.

\bibitem{ding20224bit}
S.~Ding, P.~Meadowlark, Y.~He, L.~Lew, S.~Agrawal, and O.~Rybakov, ``4-bit conformer with native quantization aware training for speech recognition,'' in \emph{Proc. Interspeech}, 2022, pp. 1711--1715.

\bibitem{zhen2022sub8bit}
K.~Zhen, H.~D. Nguyen, R.~Chinta, N.~Susanj, A.~Mouchtaris, T.~Afzal, and A.~Rastrow, ``Sub-8-bit quantization aware training for 8-bit neural network accelerator with on-device speech recognition,'' in \emph{Proc. Interspeech}, 2022, pp. 3033--3037.

\bibitem{wang2023bitnet}
{H. Wang}, {S. Ma}, {L. Dong}, {S. Huang}, {H. Wang}, {L. Ma}, {F. Yang}, {R. Wang}, {Y. Wu}, and {F. Wei}, ``Bitnet: Scaling 1-bit transformers for large language models,'' \emph{arXiv preprint arXiv:2310.11453}, 2023.

\bibitem{ma2024era}
S.~Ma, H.~Wang, L.~Ma, L.~Wang, W.~Wang, S.~Huang, L.~Dong, R.~Wang, J.~Xue, and F.~Wei, ``The era of 1-bit llms: All large language models are in 1.58 bits,'' \emph{arXiv preprint arXiv:2402.17764}, 2024.

\bibitem{xnor2016rastegari}
M.~Rastegari, V.~Ordonez, J.~Redmon, and A.~Farhadi, ``Xnor-net: Imagenet classification using binary convolutional neural networks,'' in \emph{Proc. ECCV}, 2016, pp. 525--542.

\bibitem{Bulat2019XNORNet}
A.~Bulat and G.~Tzimiropoulos, ``Xnor-net++: Improved binary neural networks,'' in \emph{Proc. BMVC}, 2019.

\bibitem{QIN2020107281}
{H. Qin}, {R. Gong}, {X. Liu}, {X. Bai}, {J. Song}, and {N. Sebe}, ``Binary neural networks: A survey,'' \emph{Pattern Recognition}, vol. 105, p. 107281, 2020.

\bibitem{Courbariaux2016binary}
{M. Courbariaux}, {I. Hubara}, {D. Soudry}, {R. El-Yaniv}, and {Y. Bengio}, ``Binarized neural networks: Training deep neural networks with weights and activations constrained to +1 or -1,'' \emph{arXiv preprint arXiv:1602.02830}, 2016.

\bibitem{Xia2023basic}
{B. Xia}, {Y. Zhang}, {Y. Wang}, {Y. Tian}, {W. Yang}, {R. Timofte}, and {L. Van Gool}, ``Basic binary convolution unit for binarized image restoration network,'' in \emph{Proc. ICLR}, 2023.

\bibitem{rybakov20232bit}
{O. Rybakov}, {P. Meadowlark}, {S. Ding}, {D. Qiu}, {J. Li}, {D. Rim}, and {Y. He}, ``2-bit conformer quantization for automatic speech recognition,'' in \emph{Proc. Interspeech}, 2023, pp. 4908--4912.

\bibitem{layernorm2016ba}
{J. L. Ba}, {J. R. Kiros}, and {G. E. Hinton}, ``Layer normalization,'' \emph{arXiv:1607.06450}, 2016.

\bibitem{Bengio2013ste}
{Y. Bengio}, {N. L\'eonard}, and {A. Courville}, ``Estimating or propagating gradients through stochastic neurons for conditional computation,'' \emph{arXiv preprint arXiv:1308.3432}, 2013.

\bibitem{koizumi2023librittsr}
{Y. Koizumi}, {H. Zen}, {S. Karita}, {Y. Ding}, {K. Yatabe}, {N. Morioka}, {M. Bacchiani}, {Y. Zhang}, {W. Han}, and {A. Bapna}, ``{LibriTTS-R: A restored multi-speaker text-to-speech corpus},'' in \emph{Proc. Interspeech}, 2023, pp. 5496--5500.

\bibitem{mcauliffe2017montreal}
M.~McAuliffe, M.~Socolof, S.~Mihuc, M.~Wagner, and M.~Sonderegger, ``Montreal forced aligner: Trainable text-speech alignment using kaldi.'' in \emph{Proc. Interspeech}, vol. 2017, 2017, pp. 498--502.

\bibitem{lim2022jets}
{D. Lim}, {S. Jung}, and {E. Kim}, ``{JETS}: Jointly training {FastSpeech2} and {HiFi-GAN} for end to end text to speech,'' in \emph{Proc. Interspeech}, 2022, pp. 21--25.

\bibitem{vaswani2017attention}
A.~Vaswani, N.~Shazeer, N.~Parmar, J.~Uszkoreit, L.~Jones, A.~N. Gomez, {\L}.~Kaiser, and I.~Polosukhin, ``Attention is all you need,'' in \emph{Proc. NeurIPS}, 2017, pp. 5998--6008.

\bibitem{kong2020hifi}
{J. Kong}, {J. Kim}, and {J. Bae}, ``{HiFi-GAN}: Generative adversarial networks for efficient and high fidelity speech synthesis,'' in \emph{Proc. NeurIPS}, vol.~33, 2020, pp. 17\,022--17\,033.

\bibitem{Bishop94mixturedensity}
C.~M. Bishop, ``Mixture density networks,'' \emph{Aston University, Birmingham UK}, 1994.

\bibitem{du2021phone}
C.~Du and K.~Yu, ``Phone-level prosody modelling with {GMM}-based {MDN} for diverse and controllable speech synthesis,'' \emph{IEEE/ACM Trans. on Audio, Speech, and Lang. Process.}, vol.~30, pp. 190--201, 2022.

\bibitem{Ilya2019adamw}
{I. Loshchilov} and {F. Hutter}, ``Decoupled weight decay regularization,'' in \emph{Proc. ICLR}, 2019.

\bibitem{Li2019explr}
{Z. Li} and {S. Arora}, ``An exponential learning rate schedule for deep learning,'' \emph{arXiv preprint arXiv:1910.07454}, 2019.

\bibitem{hayashi2020esp}
T.~Hayashi, R.~Yamamoto, K.~Inoue, T.~Yoshimura, S.~Watanabe, T.~Toda, K.~Takeda, Y.~Zhang, and X.~Tan, ``{ESPnet-TTS}: Unified, reproducible, and integratable open source end-to-end text-to-speech toolkit,'' in \emph{Proc. ICASSP}, 2020, pp. 7654--7658.

\bibitem{huffman52}
D.~A. Huffman, ``A method for the construction of minimum-redundancy codes,'' \emph{Proc. IRE}, vol.~40, no.~9, pp. 1098--1101, 1952.

\end{thebibliography}

\end{document}